\documentclass[twocolumn,showpacs,preprintnumbers,amsmath,amssymb,prl]{revtex4}

\usepackage{graphicx}
\usepackage{dcolumn}
\usepackage{bm}
\usepackage{color}
\usepackage{notes2bib}

\begin{document}

\preprint{}

\title{High precision vector magnetometry with uniaxial quantum centers in silicon carbide}

\author{D.~Simin$^{1}$}
\author{F.~Fuchs$^{1}$}
\author{H.~Kraus$^{1}$}
\author{A.~Sperlich$^{1}$}
\author{P.~G.~Baranov$^{2}$}
\author{G.~V.~Astakhov$^{1}$}
\email[E-mail:~]{astakhov@physik.uni-wuerzburg.de}
\author{V.~Dyakonov$^{1,3}$}
\email[E-mail:~]{dyakonov@physik.uni-wuerzburg.de}

\affiliation{$^1$Experimental Physics VI, Julius-Maximilian University of W\"{u}rzburg, 97074 W\"{u}rzburg, Germany \\
$^2$Ioffe Physical-Technical Institute, 194021 St.~Petersburg, Russia\\ 
$^3$Bavarian Center for Applied Energy Research (ZAE Bayern), 97074 W\"{u}rzburg, Germany}

\begin{abstract}
We show that uniaxial color centers in silicon carbide with hexagonal lattice structure can be used to measure not only the strength but also the polar angle of the external magnetic field with respect to the defect axis with high precision. The method is based on the optical detection of multiple spin resonances in the silicon vacancy defect with quadruplet ground state. We achieve a perfect agreement between the experimental and calculated spin resonance spectra without any fitting parameters, providing angle resolution of a few degrees in the magnetic field range up to several millitesla. Our approach is suitable for ensembles as well as for single \mbox{spin-3/2} color centers, allowing for vector magnetometry on the nanoscale at ambient conditions. 
\end{abstract}

\date{\today}


\maketitle

Optically addressable atomic-scale spin centers constitute the basis for nanomagnetometry with high sensitivity and high spatial resolution \cite{Chernobrod:2005jb,Taylor:2008cp}. The most prominent example is the nitrogen-vacancy (NV) defect in diamond and several benchmark experiments have been performed using this system \cite{Maze:2008cs, Balasubramanian:2008cz, Degen:2008ih}, including proton nuclear magnetic resonance on the nanometer scale \cite{Mamin:2013eu, Staudacher:2013kn}. The principle of magnetometry with spin-carrying color centers is based on optical detection of magnetic resonance (ODMR), subject to external magnetic field. In case of individual NV defects with spin $S = 1$, the projection of the magnetic field on the defect axis is measured. The NV defect in the diamond cubic lattice is oriented along one out of four $\langle 111 \rangle$ crystallographic axes and, therefore, using ensemble experiments the magnetic field vector $\mathbf{B}$ can be reconstructed \cite{Maertz:2010ko, Steinert:2010kk}. Ensembles of the NV defects are also suggested for the implementation of high precision magnetic field sensors with femtotesla sensitivity  \cite{Acosta:2009gu, Wolf:2014us} and solid-state frequency standards \cite{Hodges:2013cs}. These implementations require high homogeneity of the NV centers. The NV defects can be fabricated with preferential alignment \cite{Michl:2014fm, Lesik:2014jp}, and using nonlinear shift of the ODMR lines in relatively high magnetic fields of several tens of millitesla the transverse field component can be reconstructed  \cite{Tetienne:2012fv}. However, in many demanding applications much lower magnetic fields should be detected, and the information on the magnetic field orientation is difficult to extract in this approach. 

\begin{figure}[t]
\includegraphics[width=.47\textwidth]{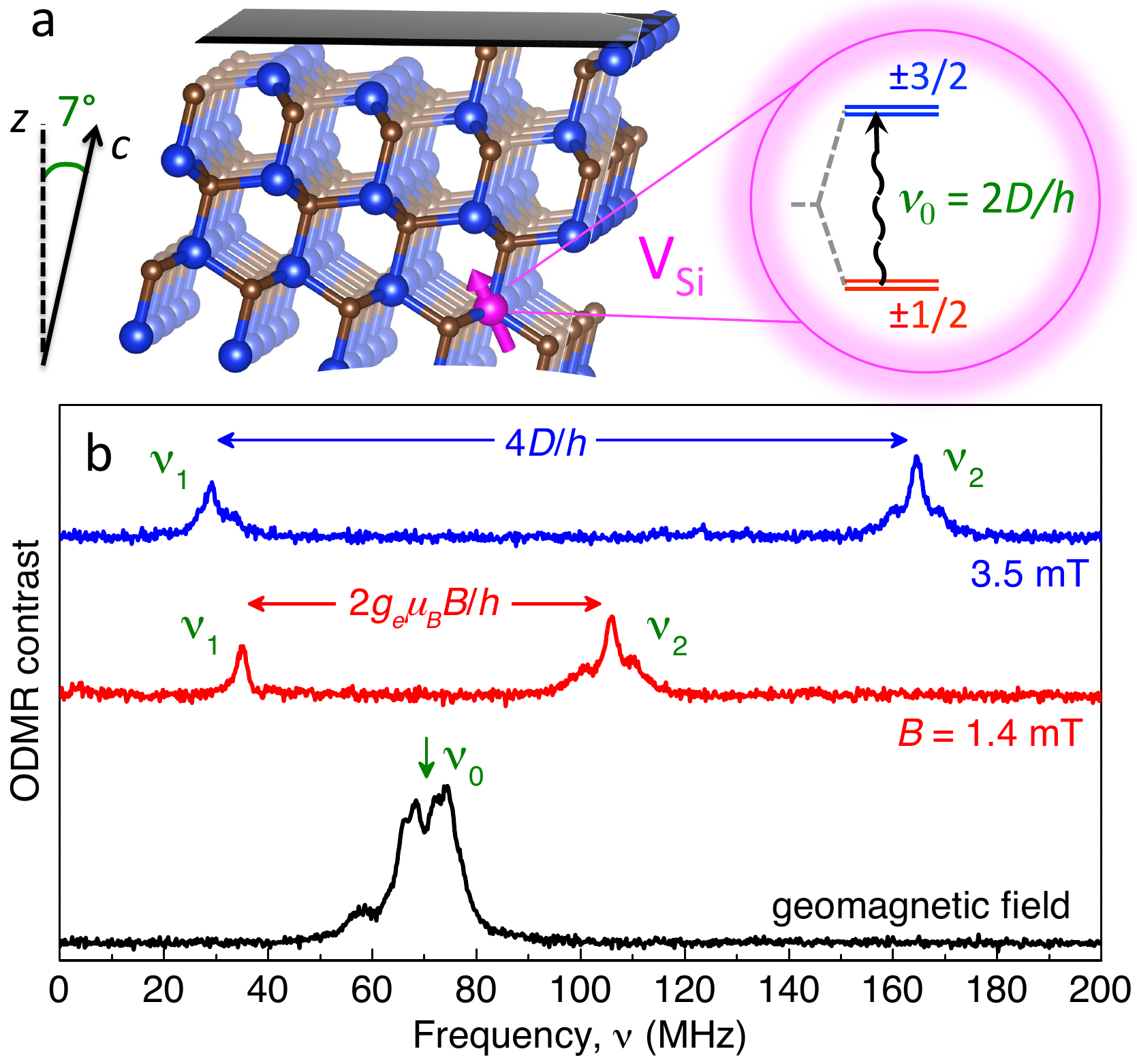}
\caption{ (a)  The $c$-axis of the 4H-SiC crystal is oriented at an angle of $7^{\circ}$ with respect to the surface normal. The silicon vacancy $\mathrm{V_{Si} (V2)}$  has spin-3/2 ground state with zero-field splitting $\nu_{0}  =  70 \, \mathrm{MHz}$. (b) Room-temperature ODMR spectra of the $\mathrm{V_{Si} (V2)}$ defect in different magnetic fields applied parallel to the surface normal ($B \| z$). } \label{fig1}
\end{figure}

Here, we demonstrate an alternative approach to implement vector magnetometry for magnetic fields below several millitesla, which is suitable for ensemble as well as for individual uniaxial spin centers with $S = 3 /2$ \cite{Kraus:2013di}.  As a model system, we consider a silicon vacancy ($\mathrm{V_{Si}}$) in silicon carbide (SiC) \cite{Vainer:1981vj, Sorman:2000ij, Mizuochi:2002kl, Baranov:2011ib}. Due to the polymorphism of SiC, there is a large variety of vacancy-related defects with appealing quantum properties \cite{Koehl:2011fv, Soltamov:2012ey, Riedel:2012jq, Fuchs:2013dz, Castelletto:2013el, Falk:2013jq, Kraus:2013di, Kraus:2013vf, Klimov:2013ua, Falk:2014fh, Christle:2014ti, Widmann:2014ve, Fuchs:2014tc, Muzha:2014th}.  All experiments presented here have been performed at room temperature on a 4H-SiC bulk crystal, possessing hexagonal lattice structure. The crystal has been grown by the standard sublimation technique, such that the  $\left[ 0001 \right]$ crystallographic direction ($c$-axis) is inclined at an angle of $7^{\circ}$ to the surface normal, i.e., to the $z$-axis of the laboratory coordinate system [Fig.~\ref{fig1}(a)]. The crystal has been irradiated with neutrons ($5 \, \mathrm{MeV}$) with a fluence of $10^{16} \, \mathrm{cm^{-2}}$, in order to generate $\mathrm{V_{Si}}$ defects. Their presence is verified by the characteristic photoluminescence (PL) in the near infrared spectral range \cite{Hain:2014tl}. 

The $\mathrm{V_{Si}}$ defects in hexagonal SiC have spin-3/2 ground state, which is split in two spin sublevels $m_S = \pm 1/ 2$ and  $m_S = \pm 3/ 2$ at zero magnetic field  \cite{Kraus:2013vf}. The zero-field splitting $2 D$ between these sublevels depends on the lattice site and polytype. In 4H-SiC there are two nonequivalent lattice cites and hence two different $\mathrm{V_{Si}}$ defects. They are distinguished by their spectroscopic features and labeled as V1 and V2 centers  \cite{Sorman:2000ij}. Here, we present results for the $\mathrm{V_{Si} (V2)}$ center with $ 2 D / h  =  70 \, \mathrm{MHz}$ [Fig.~\ref{fig1}(a)]. A detailed  characterization of the system is presented elsewhere \cite{Riedel:2012jq, Kraus:2013di, Kraus:2013vf, Hain:2014tl, Widmann:2014ve, Fuchs:2014tc}

A diode laser operating at $785 \, \mathrm{nm}$ is used to optically pump the $\mathrm{V_{Si} (V2)}$ center in the host SiC with a bandgap of $3.23 \, \mathrm{eV}$. The optical excitation followed by spin-dependent recombination results in preferential population of the $m_S = \pm 1/ 2$ sublevels \bibnote{Whether the $m_S = \pm 1/ 2$ or $m_S = \pm 3/ 2$ sublevels are preferentially populated varies with polytype and crystallographic site.}. The PL from these centers is passed through 800-nm and  850-nm longpass filters and detected by a Si photodiode (up to $1050 \, \mathrm{nm}$). The PL intensity is spin dependent, in case of $\mathrm{V_{Si} (V2)}$ it is higher when the system is initialized into the $m_S = \pm 3/ 2$ states. The radiofrequency (RF) provided by a signal generator (2~dBm power) is guided to a thin copper wire  and terminated with 50~$\Omega$  impedance. The laser beam is focused close to the wire using a 10$\times$ optical objective ($\mathrm{N.A. }= 0.25$) and the PL is collected through the sample using a biconvex lens.  The $x$-axis of the laboratory coordinate system is set parallel to the wire. 
The external magnetic field $\mathbf{B}$ 
can be applied along arbitrary directions, using a 3D coil arrangement. The magnet is calibrated using a 3D Hall sensor, providing angle resolution of $2^{\circ}$. 

Without external magnetic field, a resonance RF $\nu_0 = 2 D / h$  induces magnetic dipole transitions between the spin-split sublevels $(\pm1/2  \rightarrow \pm3/2)$ [Fig.~\ref{fig1}(a)], resulting in a change of the PL intensity ($\mathrm{\Delta PL}$). To increase the sensitivity, the RF is chopped at  $677 \, \mathrm{Hz}$ and the output PL signal is locked-in. An example of the ODMR spectrum (i.e., the ODMR contrast $\mathrm{\Delta PL / PL}$ against applied RF frequency) obtained for $B \rightarrow 0$ is shown in  Fig.~\ref{fig1}(b) (lower curve). The ODMR line is split around $\nu_0 = 70 \, \mathrm{MHz}$ due to the geomagnetic field and stray magnetic fields.  

We now discuss the evolution of the ODMR spectra in external magnetic fields. The corresponding spin Hamiltonian is written in the form
\begin{equation}
\mathcal{H} = g_e \mu_B   \mathbf{B}  \mathbf{S} + D[ S_c^2 - S(S+1) / 3] \,.
 \label{Hamiltonian}
\end{equation}

Here, $g_e \approx 2.0$ is the electron g-factor, $\mu_B$ is the Bohr magneton and $S_c$ is the projection of the total spin on the $c$-axis of 4H-SiC \cite{Kraus:2013vf}. For the sake of simplicity, we do not consider the hyperfine interaction. We also neglect any deviation from the uniaxial symmetry, described by the transverse zero-field splitting parameter $E \ll D$. Remarkably, the $m_S = \pm 1/ 2$ ($m_S = \pm 3/ 2$) states remain doubly degenerated even in the presence of electric and strain fields, in accord with the Kramers theorem. We hence take $E = 0$ as a good approximation \cite{Kraus:2013vf, Widmann:2014ve}.   

\begin{figure}[t]
\includegraphics[width=.47\textwidth]{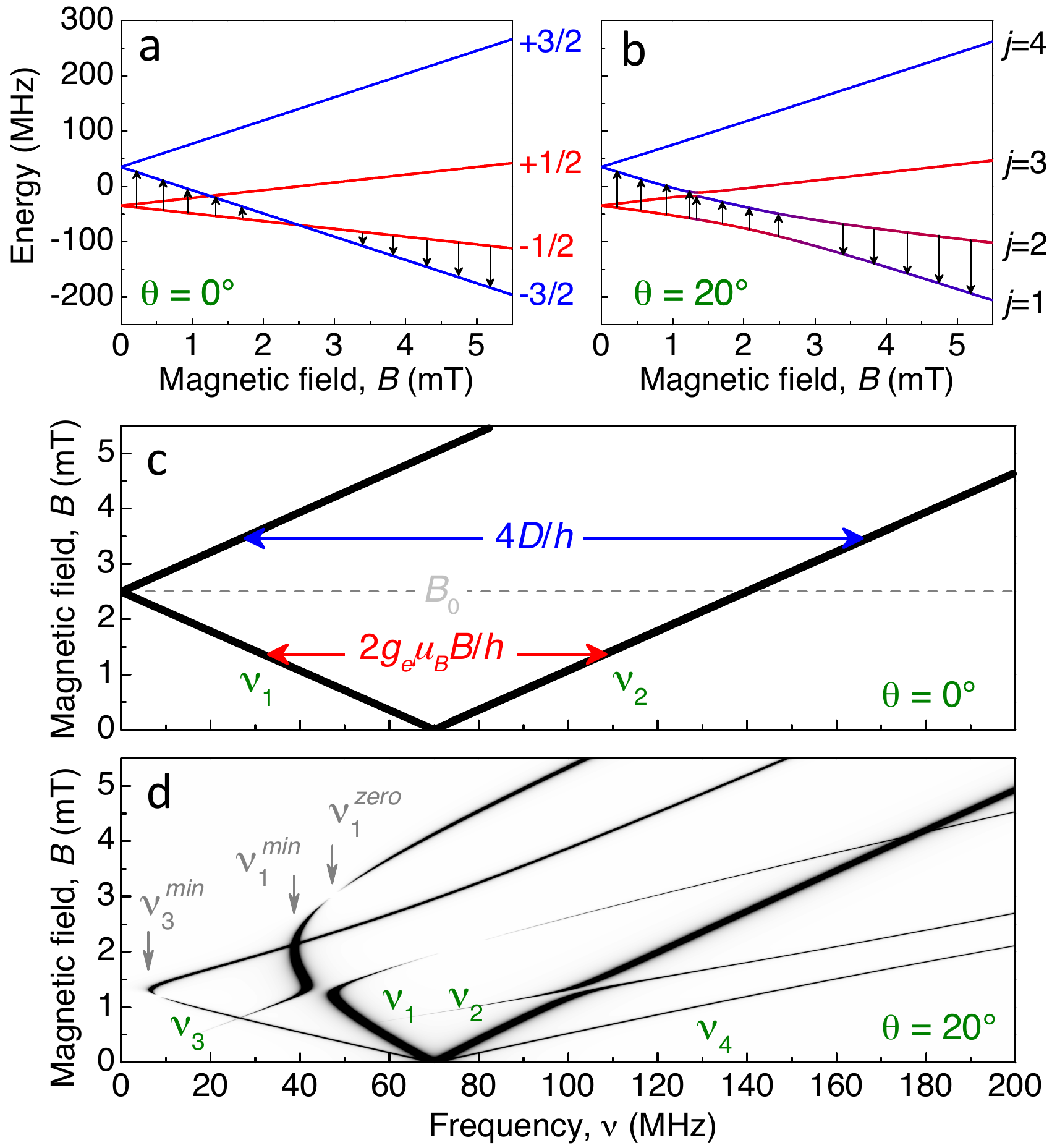}
\caption{ Spin sublevels in the $\mathrm{V_{Si} (V2)}$  ground state, calculated using Hamiltonian~(\ref{Hamiltonian}) as a function of the magnetic field $B$ applied (a) parallel to the $c$-axis and (b) at an angle $\theta = 20^{\circ}$ with respect to the $c$-axis. Magnetic field evolution of the ODMR spectrum calculated for (c) $\theta = 0^{\circ}$ and (d) $\theta = 20^{\circ}$. The ODMR contrast is coded by the line thickness. The transitions, associated with the $\nu_1$ ODMR line, are shown by arrows in (a) and (b).} \label{fig2}
\end{figure}

When external magnetic field is applied parallel to the symmetry axis $B \| c$ ($\theta = 0^{\circ}$), the spin states are split as $\epsilon_{\pm 1/2} = -D \pm g_e \mu_B B /2$ and $\epsilon_{\pm 3/2} = +D \pm 3 g_e \mu_B B /2$ [Fig.~\ref{fig2}(a)]. One of the dipole-allowed transition with $\Delta m_S = - 1$ is $(-1/2 \rightarrow -3/2)$ and the corresponding ODMR line shifts linearly with magnetic field $\nu_1 = | \nu_0 - g_e   \mu_B B / h |$, as shown in Fig.~\ref{fig2}(c). Another dipole-allowed transition with $\Delta m_S = + 1$ is $(+1/2  \rightarrow +3/2)$ and the corresponding ODMR line shifts linearly towards higher frequencies $\nu_2 =  \nu_0 + g_e   \mu_B B / h$.  For magnetic fields lower than $B_0 = h \nu_0 / (g_e   \mu_B) = 2.5 \, \mathrm{mT}$ the splitting between the ODMR lines yields $\nu_2 - \nu_1 = 2 g_e   \mu_B B / h$ [middle curve in Fig.~\ref{fig1}(b)] and for higher magnetic fields $B > B_0$ this splitting is equal $\nu_2 - \nu_1 = 4 D / h$ [upper curve in Fig.~\ref{fig1}(b)]. Remarkably, the dipole-allowed transition $(-1/2  \rightarrow +1/2)$, expected between $\nu_1$ and $\nu_2$ at frequencies $g_e   \mu_B B / h$, is not observed in the ODMR spectra. The reason is the equal population of the $m_S = \pm 1/ 2$ sublevels due to the optical pumping. 

The behavior becomes much more complex when the magnetic field vector $\mathbf{B}$ is no more parallel to the defect symmetry axis. It is instructive to consider a certain pronounced  case in detail, namely when the magnetic field is applied at an angle $\theta = 20^{\circ}$ with respect to the $c$-axis. By solving Hamiltonian~(\ref{Hamiltonian}), we find the energy level positions and the eigenfunctions $| j \rangle$  ($j=1,2,3,4$) depending on $B$ [Fig.~\ref{fig2}(b)]. The mixing of the $m_S = \pm 1/ 2$ states in the transverse component of the magnetic field $B \sin \theta$ results in the appearance of two additional ODMR lines, which are labeled $\nu_3$ and  $\nu_4$ in Fig.~\ref{fig2}(d). The level anticrossing seen in Fig.~\ref{fig2}(b) at $1.3 \, \mathrm{mT}$ (between states $| 2 \rangle$ and $| 3 \rangle$) and at $2.0 \, \mathrm{mT}$ (between states $| 1\rangle$ and $| 2 \rangle$) reveals as the appearance of 'turning points' at frequencies $\nu_{3}^{min} = 6 \, \mathrm{MHz}$ and $\nu_{1}^{min} = 39 \, \mathrm{MHz}$ in the calculated spectra of Fig.~\ref{fig2}(d). Furthermore, according to Fig.~\ref{fig2}(b) the population of states $| 1 \rangle$ and $| 2 \rangle$  becomes equal at $3.0 \, \mathrm{mT}$ and the ODMR contrast  tends to zero, seeing as a discontinuity at $\nu_{1}^{zero} = 47 \, \mathrm{MHz}$ in Fig.~\ref{fig2}(d).  

\begin{figure}[t]
\includegraphics[width=.47\textwidth]{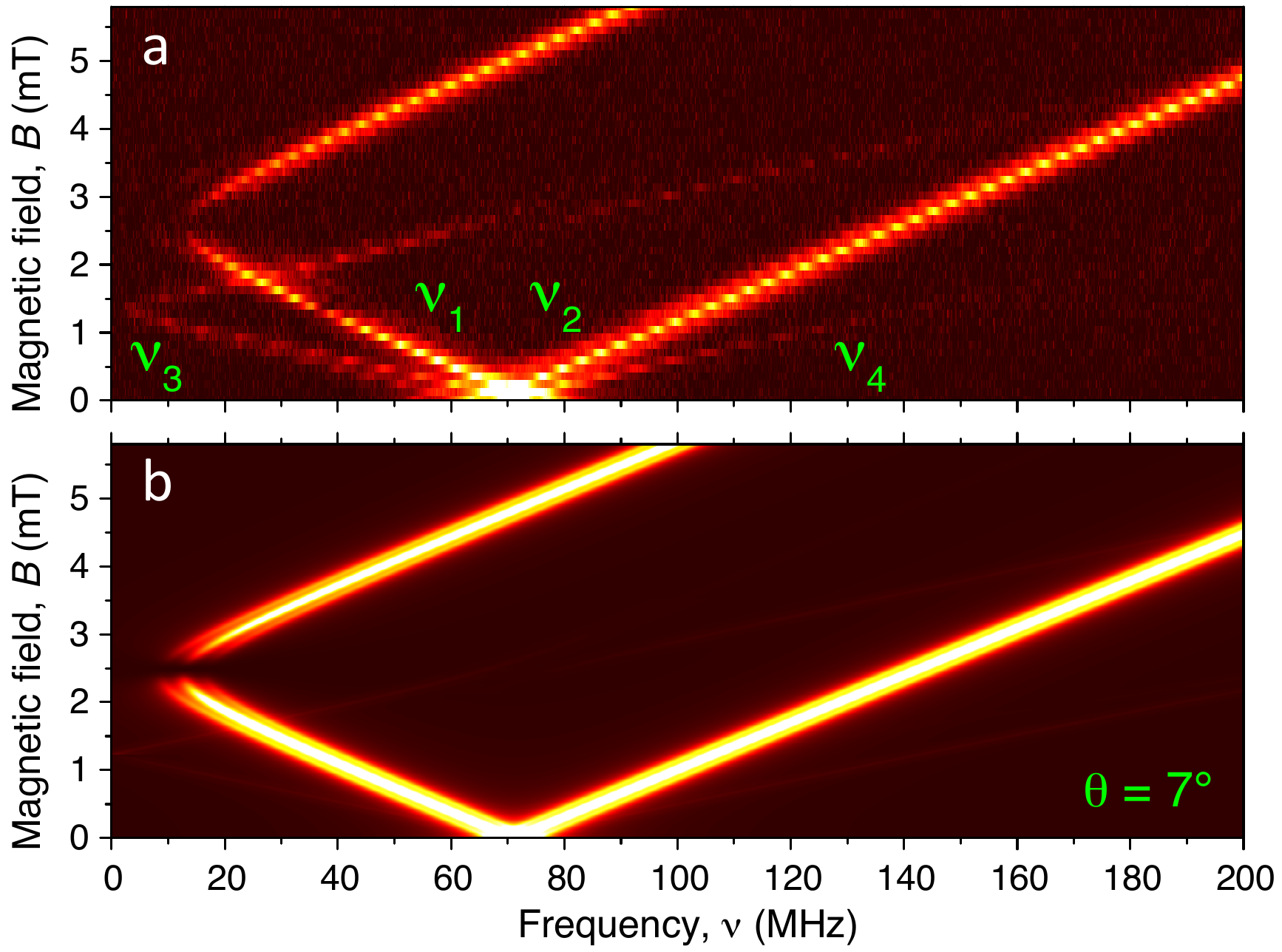}
\caption{ Evolution of the $\mathrm{V_{Si} (V2)}$ ODMR spectra in magnetic fields applied parallel to the surface normal ($\mathbf{B} \| z$), i.e., at an angle $\theta = 7^{\circ}$ with respect to the $c$-axis: (a) experiment and (b) calculation. The ODMR contrast is color-coded (bright colors correspond to higher values). } \label{fig3}
\end{figure}

We now explain how the field-dependent ODMR spectra in Fig.~\ref{fig2}(d) are calculated. To obtain the probability of the RF-induced transitions between states $| j \rangle$ and  $| k \rangle$, we apply 
\begin{equation}
W_{jk} \sim  \big| \langle j | \mathbf{B_1 S_1} | k \rangle \big|^2\,,
 \label{Transitions}
\end{equation}
Here, $\mathbf{B_1}$ is the driving RF-field (in our experiments $\mathbf{B_1} \| y$). We modify the spin-3/2 matrices $\mathbf{S_1}$ 
in Eq.~(\ref{Transitions}) to take into account that the $m_S = \pm 1/2$ states are equally populated due to the optical pumping and hence do not contribute to the ODMR signal. Namely, the matrix elements 
coupling these states are set to zero. Finally we simulate the ODMR spectra assuming Lorentzian line shapes with the experimental value for the full width at half maximum of $2.7 \, \mathrm{MHz}$ \cite{Kraus:2013vf}.  One can see from Fig.~\ref{fig2}(d) that for an arbitrary strength and orientation of the magnetic field up to six ODMR lines can be observed.  

Using Eq.~(\ref{Transitions}) with modified spin matrices we can perfectly describe our experimental data. Figure~\ref{fig3}(a) presents the evolution of the ODMR spectra when the external magnetic field is applied parallel to the $z$-axis of the laboratory coordinate system, i.e. at an angle $\theta = 7^{\circ}$ with respect to the $c$-axis [see Fig.~\ref{fig1}(a)]. Because the deviation from the defect symmetry axis is small, the behavior is very similar to the high-symmetry case ($B \| c$) of Fig.~\ref{fig2}(c). The difference is the manifestation of anticrossing ($\nu_{1}^{min} = 13 \, \mathrm{MHz}$) between the ($m_S = -3/2$)-like and ($m_S = -1/2$)-like states at $2.5 \, \mathrm{mT}$. Furthermore, in addition to the inner resonances $\nu_{1}$ and $\nu_{2}$, the outer resonances $\nu_{3}$ and $\nu_{4}$ appear.  They shift with double slope of the inner resonances. This behavior is closely reproduced by our calculations shown in Fig.~\ref{fig3}(b). 

\begin{figure}[t]
\includegraphics[width=.47\textwidth]{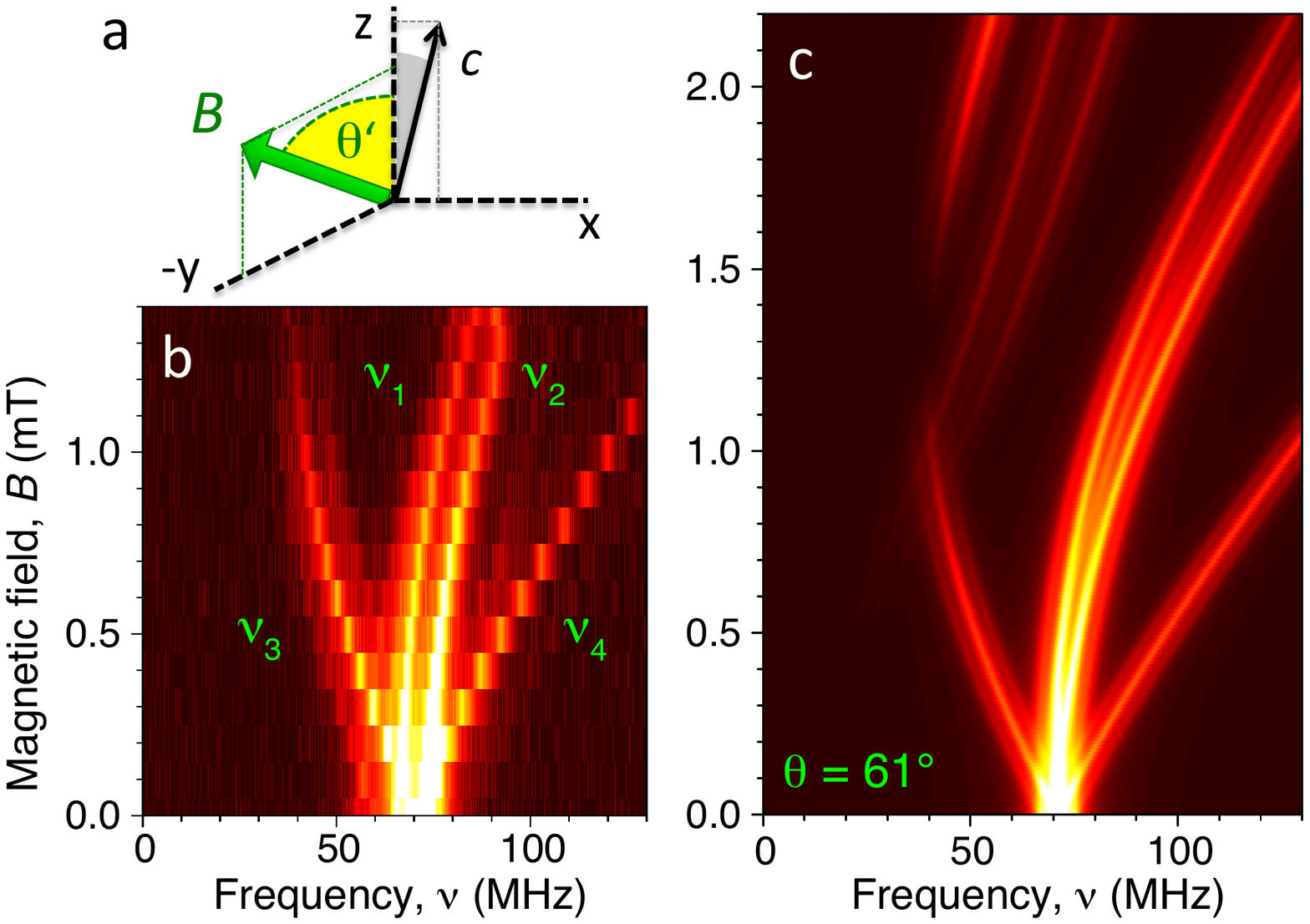}
\caption{ (a) Evolution of the $\mathrm{V_{Si} (V2)}$ ODMR spectra in magnetic fields applied at an angle $\theta' = 60^{\circ}$ with respect to the surface normal in the yz-plane: (b) experiment and (c) calculation. } \label{fig4}
\end{figure}

As another demonstration, we measure the evolution of the ODMR spectra when the external magnetic field is applied in the $yz$-plane at an angle $\theta' = 60^{\circ}$  with respect to the $z$-axis [Fig.~\ref{fig4}(a)]. This corresponds to the angle $\theta \approx 61^{\circ}$ between $\mathbf{B}$ and the $c$-axis. It is close to the magic angle $\theta_m =  \arccos (1/\sqrt{3}) \approx  54.7^{\circ}$, when the splitting between the inner ODMR resonances vanishes.  Indeed, we experimentally observed $\nu_{2} - \nu_{1}  \ll \nu_{4} - \nu_{3}$  [Fig.~\ref{fig4}(b)]. Furthermore, when the magnetic field is significantly inclined from the symmetry axis, the outer resonances $\nu_{3}$ and $\nu_{4}$ become much more pronounced. This complicated behaviour is reproduced in the calculated ODMR spectra of Fig.~\ref{fig4}(c) as well. 

\begin{figure}[t]
\includegraphics[width=.47\textwidth]{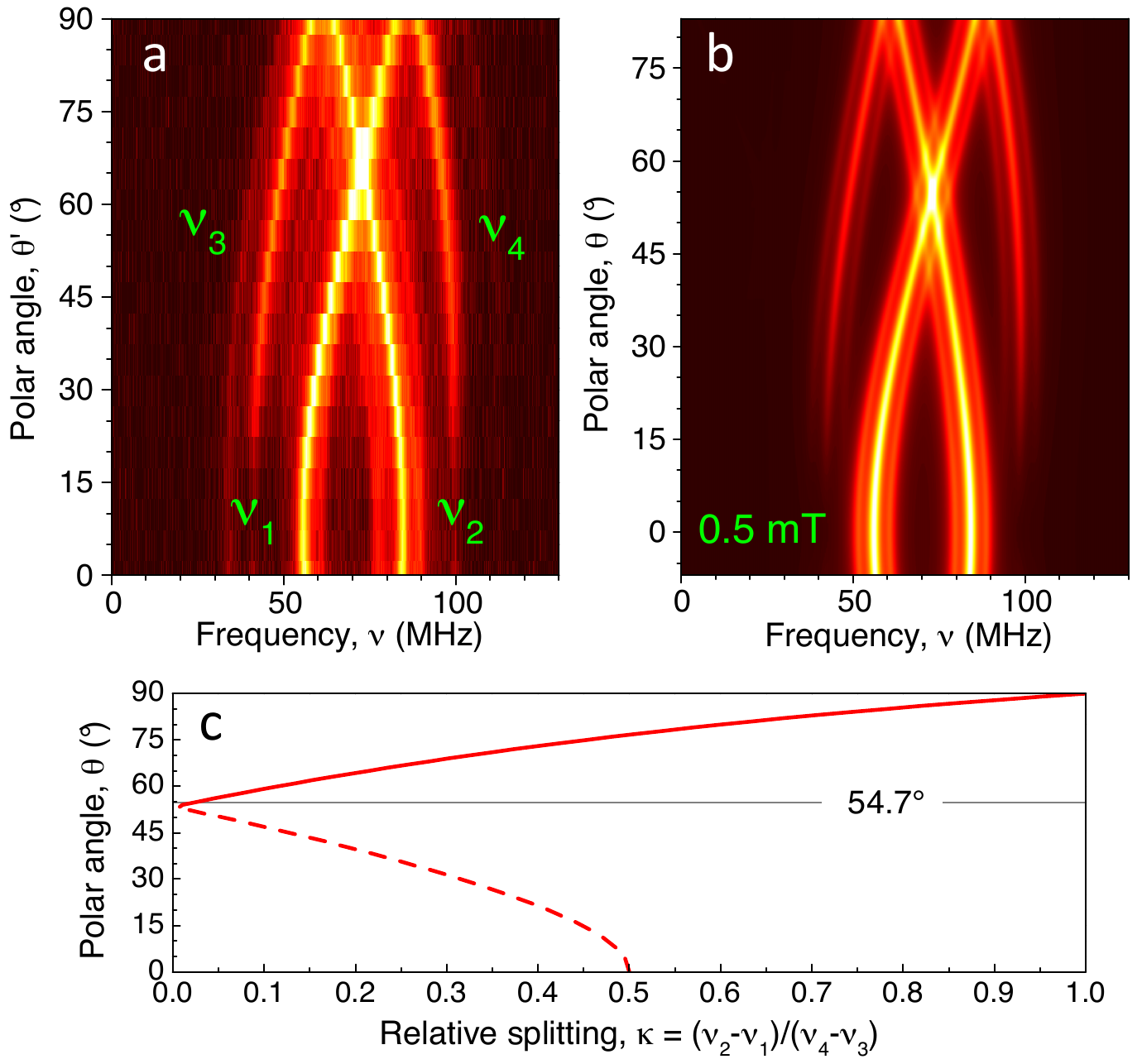}
\caption{Evolution of the $\mathrm{V_{Si} (V2)}$ ODMR spectra  in a magnetic field $B = 0.5 \, \mathrm{mT}$ revolving in the  xz-plane: (a) experiment and (b) calculation. The vertical axis in (b) is shifted by $7^{\circ}$ to account for the $c$-axis tilting with respect to the z-axis. (c)  Polar angle dependence of the relative splittings between inner ($\nu_2 - \nu_1$) and outer ($\nu_4 - \nu_3$) resonances.  The dashed line corresponds to the angles when the outer ODMR lines are weak. } \label{fig5} 
\end{figure}

We now discuss how the orientation of the magnetic field can be reconstructed. Figure~\ref{fig5}(a) shows the evolution of the ODMR spectra depending on the magnetic field orientation at a fixed strength $B = 0.5 \, \mathrm{mT}$. The orientation angle varies from $\theta' = 0^{\circ}$ ($\mathbf{B} \| z$) to $\theta' = 90^{\circ}$ ($\mathbf{B} \bot z$) in the $xz$-plane. 
As the defect symmetry axis is not exactly parallel to the $z$-axis and lies in the $xz$-plane [the schematic of Fig.~\ref{fig4}(a)] we shift the calculated results accordingly ($\theta = \theta' - 7^{\circ}$) for the direct comparison [Fig.~\ref{fig5}(b)]. As one can see from Figs.~\ref{fig5}(a) and (b), the splitting between the inner ($\nu_2 - \nu_1$) and outer ($\nu_4 - \nu_3$) resonances changes with magnetic field orientation. Figure~\ref{fig5}(c) represents the interconnection between the polar angle $\theta$, i.e., the angle between the magnetic field vector and the $c$-axis, and the relative splitting between resonances $\kappa = (\nu_2 - \nu_1) / (\nu_4 - \nu_3)$. 

Thus, the algorithm to determine the orientation of the magnetic field can be described as follows. For the relatively large polar angles $\theta > 54.7^{\circ}$ the $m_S = \pm 1/ 2$ states are strongly mixed and the outer resonances are well resolved. To determine $\theta$ in this case, one measures the relative distance between the ODMR lines $\kappa$ and uses the calibration curve shown by the solid line in Fig.~\ref{fig5}(c). In the opposite case when $\theta < 54.7^{\circ}$, the $m_S = \pm 1/ 2$ states are weakly mixed and the outer ODMR lines have lower amplitude, less than 50\% compared to the inner ODMR lines. The corresponding calibration curve to determine $\theta$ is shown by the dashed line in Fig.~\ref{fig5}(c). The angle resolution depends on the distance between the spin resonance frequencies at a given magnetic filed and the ODMR linewidth.  For the data presented in Fig.~\ref{fig5}, the anglular resolution is better than $5^{\circ}$. It should be possible to improve it further by using advanced readout protocols \cite{Taylor:2008cp}. Remarkably, at  $\theta= 0^{\circ}$ and $\theta= 90^{\circ}$ only two ODMR lines should be observed and therefore they are formally indistinguishable. However, in the presence of weak stray fields they behave differently. For  $\theta \rightarrow 0^{\circ}$ the outer resonances with double splitting become visible [as in Fig.~\ref{fig3}(a)] and, for  $\theta \rightarrow 90^{\circ}$, the outer resonances are of the same amplitude as the inner resonances and only slightly split from them. 

In summary, uniaxial \mbox{spin-3/2}  centers in hexagonal lattice can be used to measure magnetic field strength and to reconstruct magnetic field orientation with respect to the symmetry axis of the crystal (i.e., the polar angle).  The method is based on the ODMR technique and applicable for ensembles of spin centers  as well as for single centers. As a probe, we used the $\mathrm{V_{Si}}$ spin center in 4H-SiC and performed demonstrative experiments in weak magnetic fields up to $5 \, \mathrm{mT}$. The experiments are perfectly described by our model for any magnetic field orientation. It should be mentioned that this method does not provide information on the azimuthal angle, but this can potentially be overcome by using differently oriented centers, for instance along the $c$-axis and in the basal plane of the crystal. Interestingly, $\mathrm{V_{Si}}$ can be incorporated into SiC nanocrystals \cite{Muzha:2014th} and their density can be controlled over eight orders of magnitude down to single defect level \cite{Fuchs:2014tc}. The selective optical addressability \cite{Riedel:2012jq} and coherent control \cite{Soltamov:2012ey} of single $\mathrm{V_{Si}}$  centers \cite{Widmann:2014ve} have also been reported. Combination of these capabilities with our findings suggests promising perspectives for vector magnetometry and local imaging down to the nanoscale.    

This work has been supported by the German Research Foundation (DFG) under grant AS 310/4 as well as by the RFBR (14-02-91344) and by the RMES (14.604.21.0083). HK acknowledges the support of the German Academic Exchange Service (DAAD), with funds of the German Federal Ministry of Education and Research (BMBF) and EU Marie Curie Actions (DAAD P.R.I.M.E. project 57183951). 




\end{document}